\documentstyle[12pt,epsfig]{article}
\begin{document}

\begin{flushright}
{\bf hep-ph/0205032} 
\end{flushright}

\vspace{0.2cm}

\begin{center}
{\large\bf A Full Determination of the Neutrino Mass Spectrum \\ 
from Two-zero Textures of the Neutrino Mass Matrix}
\end{center}

\vspace{0.2cm}

\begin{center}
{\bf Zhi-zhong Xing} \footnote{Electronic address: 
xingzz@mail.ihep.ac.cn} \\
{\it CCAST (World Laboratory), P.O. Box 8730, Beijing 100080, China} \\  
{\it and Institute of High Energy Physics, Chinese Academy of Sciences, \\
P.O. Box 918 (4), Beijing 100039, China}
\footnote{Mailing address}
\end{center}

\vspace{2cm}

\begin{abstract}
We show that it is possible to fully determine the
neutrino mass spectrum from two-zero textures of the neutrino
mass matrix. As a consequence, definite predictions can be
obtained for the neutrinoless double beta decay. 
\end{abstract}

\newpage

In the flavor basis where the charged lepton mass matrix is diagonal,
a phenomenological analysis of two-zero textures and
Majorana CP-violating phases of the neutrino mass matrix has recently
been carried out by the author \cite{Xing02}. Seven of such textures were
found to be in accord with current 
experimental data on 
atmospheric \cite{ATM}, solar \cite{SUN} and
reactor \cite{CHOOZ} neutrino oscillations. A classification
of two-zero textures of the neutrino mass matrix was originally done by
Frampton, Glashow and Marfatia \cite{FGM}. 
In two follow-up works, Frampton, Oh and Yoshioka \cite{FOY} have proposed 
a model with three Higgs triplets to gain texture zeros 
of the neutrino mass matrix; and Kageyama, Kaneko, Shimoyama and 
Tanimoto \cite{Tanimoto} have incorporated those phenomenologically
acceptable textures with the seesaw mechanism. 

\vspace{0.3cm}

As an addendum to Ref. \cite{Xing02}, the present letter aims to show 
that it is actually possible to {\it fully} determine the neutrino mass 
spectrum from two-zero textures of the neutrino mass matrix. 
This important point was not observed in Ref. \cite{Xing02} and 
Refs. \cite{FGM,FOY,Tanimoto}, where only the ratios of
three neutrino masses were calculated. Given the neutrino mass spectrum,
a definite prediction for the neutrinoless double beta decay can be
made. Taking account of current experimental data, which remain rather
rough, we illustrate the typical magnitudes of neutrino masses and
that of the neutrinoless double beta decay for each of the seven two-zero
textures of the neutrino mass matrix. 

\vspace{0.5cm}

The Majorana neutrino mass matrix can be written as
\begin{equation}
M \; =\; V \left ( \matrix{
m_1 & 0 & 0 \cr
0 & m_2 & 0 \cr
0 & 0 & m_3 \cr} \right ) V^{\rm T} \; 
\end{equation}
in the flavor basis where the charged lepton mass matrix is diagonal,
where $m_i$ (for $i=1,2,3$) are physical neutrino masses
and $V$ stands for the lepton flavor mixing matrix linking the neutrino
mass eigenstates $(\nu_1, \nu_2, \nu_3)$ to the neutrino flavor
eigenstates $(\nu_e, \nu_\mu, \nu_\tau)$. Following
Ref. \cite{Xing02}, we parametrize $V$ as
\begin{equation}
V \; = \; \left ( \matrix{
c_x c_z & s_x c_z & s_z \cr
- c_x s_y s_z - s_x c_y e^{-i\delta} &
- s_x s_y s_z + c_x c_y e^{-i\delta} &
s_y c_z \cr 
- c_x c_y s_z + s_x s_y e^{-i\delta} & 
- s_x c_y s_z - c_x s_y e^{-i\delta} & 
c_y c_z \cr } \right ) 
\left ( \matrix{
e^{i\rho}	& 0	& 0 \cr
0	& e^{i\sigma}	& 0 \cr
0	& 0	& 1 \cr} \right ) \; ,
\end{equation}
where $s_x \equiv \sin\theta_x$, $c_y \equiv \cos\theta_y$, and so on.
In this ``standard'' parametrization of $V$, the Dirac phase $\delta$ 
controls the strength of CP violation in normal neutrino oscillations, 
while the Majorana phases $\rho$ and $\sigma$ are relevant to the neutrinoless 
double beta decay \cite{FX01}. 
Without loss of generality, three mixing angles 
($\theta_x, \theta_y, \theta_z$) can all be arranged to lie in the first
quadrant. Arbitrary values are possible for three CP-violating phases 
($\delta, \rho, \sigma$). Note that $M$ is symmetric, 
thus it consists totally of six independent complex entries. 
Assuming two of them to vanish, one may figure out fifteen distinct 
textures of $M$. Only seven textures, as listed in Table 1, 
were found to be compatible with current experimental data and empirical 
hypotheses \cite{Xing02,FGM}.
Their predictions for the ratios of three neutrino masses, 
\begin{equation}
\xi \; \equiv \; \frac{m_1}{m_3} \; , ~~~~~
\zeta \; \equiv \; \frac{m_2}{m_3} \; , 
\end{equation}
are quoted from Ref. \cite{Xing02} and listed in the same table. 
Our present purpose is to determine $m_1$, $m_2$ and $m_3$ separately,
so as to fully fix the neutrino mass spectrum.

\vspace{0.3cm}

Given $\xi$ and $\zeta$, the absolute values of three neutrino masses
can be extracted from the measured mass-squared differences of solar
and atmospheric neutrino oscillations (i.e., $\Delta m^2_{\rm sun}$
and $\Delta m^2_{\rm atm}$). Because current Super-Kamiokande 
and SNO data strongly support the hypothesis that solar and
atmospheric neutrino oscillations are dominated respectively by 
$\nu_e \rightarrow \nu_\mu$ and $\nu_\mu \rightarrow \nu_\tau$ 
transitions \cite{ATM,SUN}, 
one may simply define $\Delta m^2_{\rm sun}$ and $\Delta m^2_{\rm atm}$ as
\begin{eqnarray}
\Delta m^2_{\rm sun} & \equiv & \left | m^2_2 ~ - ~ m^2_1 \right | \; ,
\nonumber \\
\Delta m^2_{\rm atm} & \equiv & \left | m^2_3 ~ - ~ m^2_2 \right | \; .
\end{eqnarray}
In Refs. \cite{Xing02,FGM}, the large-angle MSW solution to the 
solar neutrino problem has been taken into account.
Thus $R_\nu \equiv \Delta m^2_{\rm sun}/\Delta m^2_{\rm atm} \sim {\cal O}
(10^{-2})$ has been used as a crucial criterion to single out the
textures of $M$ which are phenomenologically favored.
From Eqs. (3) and (4), it is straightforward to obtain
$$
m_3 \; = \; \sqrt{\frac{\Delta m^2_{\rm atm}}{|\zeta^2 -1|}} \;\; ,
\eqno{\rm (5a)}
$$
or equivalently
$$
m_3 \; = \; \sqrt{\frac{\Delta m^2_{\rm sun}}{|\xi^2 - \zeta^2|}} \;\; .
\eqno{\rm (5b)}
$$
Once $m_3$ is determined from Eq. (5a) or (5b), the magnitudes of
$m_1$ and $m_2$ can be fixed by use of Eq. (3) or with the help of the
formulas presented in Ref. \cite{X02}.

\vspace{0.3cm}

To be specific, let us calculate $m_3$ for each of the seven
textures of $M$ listed in Table 1. We use
$\Delta m^2_{\rm atm} \approx 3\times 10^{-3} ~ {\rm eV}^2$ \cite{ATM} and
$\Delta m^2_{\rm sun} \approx 5\times 10^{-5} ~ {\rm eV}^2$ \cite{Bahcall}
as typical inputs.
Note that the number of $\Delta m^2_{\rm sun}$ results from the
best global fit of present solar neutrino oscillation 
data (including the latest
SNO neutral current measurement \cite{SUN}) in the large-angle 
MSW mechanism, and the corresponding value of $\theta_x$ 
is $\theta_x \sim 30^\circ$. Furthermore, $\theta_y \sim 40^\circ$
and $\theta_z \sim 5^\circ$ are typically taken, although the latter
may be much smaller. We assume the unknown CP-violating phase $\delta$ 
to be around $90^\circ$ in most cases. 
As the uncertainty associated with 
$\Delta m^2_{\rm atm}$ is expected to be smaller than that
associated with $\Delta m^2_{\rm sun}$, we choose to 
use Eq. (5a) instead of Eq. (5b) in the calculation of $m_3$.

\vspace{0.3cm}

\underline{Pattern $\rm A_1:$} ~ Because of $s_z \ll 1$ and
$t_x \sim t_y$, $\zeta^2 \ll 1$ is naturally anticipated. Therefore we obtain
\setcounter{equation}{5}
\begin{equation}
m_3 \; \approx \; \sqrt{\Delta m^2_{\rm atm}} \; \approx \; 
5.5 \times 10^{-2} ~ {\rm eV} \; .
\end{equation}
Using the typical inputs $\theta_x = 30^\circ$,
$\theta_y = 40^\circ$ and $\theta_z = 5^\circ$, we arrive at
\begin{equation}
m_1 \; \approx \; 2.3 \times 10^{-3} ~ {\rm eV} \; , ~~~
m_2 \; \approx \; 7.0 \times 10^{-3} ~ {\rm eV} \; .
\end{equation}
This quasi-hierarchical spectrum of neutrino masses implies that
it is in practice impossible to detect the neutrinoless double beta decay.

\vspace{0.3cm}

\underline{Pattern $\rm A_2$:} ~ Once again $\zeta^2 \ll 1$ holds. 
Then we obtain the same value of $m_3$ as that given in Eq. (6).
Using the same inputs as above, we find
\begin{equation}
m_1 \; \approx \; 3.3 \times 10^{-3} ~ {\rm eV} \; , ~~~
m_2 \; \approx \; 1.0 \times 10^{-2} ~ {\rm eV} \; .
\end{equation}
One can see that the neutrino mass spectra of Patterns $\rm A_1$
and $\rm A_2$ are quite similar.

\vspace{0.3cm}

\underline{Pattern $\rm B_1$:} ~ With the help of Eq. (5a), we obtain
\begin{equation}
m_3 \; \approx \; \sqrt{\frac{\Delta m^2_{\rm atm}}{|t^4_y -1|}} \; \approx \; 
7.7 \times 10^{-2} ~ {\rm eV} \; ,
\end{equation}
where $\theta_y = 40^\circ$ has been used. In the lowest-order approximation,
we have
\begin{equation}
m_1 \; \approx \; m_2 \; \approx \; 5.4 \times 10^{-2} ~ {\rm eV} \; .
\end{equation}
This quasi-degenerate spectrum of neutrino masses might have useful hints
at possible flavor symmetries and their breaking schemes, which are 
expected to be responsible for the generation of lepton masses \cite {FX96} 
and associated with the origin of leptogenesis \cite{Yanagida}.
It is also worth mentioning that the values of $m_i$ in Eqs. (9) and (10) are
compatible with the present direct-mass-search experiments \cite{PDG},
in particular for the electron neutrino. 

\vspace{0.3cm}
 
\underline{Pattern $\rm B_2$:} ~ In analogy to Eq. (9), $m_3$ reads as 
\begin{equation}
m_3 \; \approx \; t^2_y \sqrt{\frac{\Delta m^2_{\rm atm}}{|t^4_y -1|}} 
\; \approx \; 5.4 \times 10^{-2} ~ {\rm eV} \; ,
\end{equation}
where $\theta_y = 40^\circ$ has been used. To lowest order, we obtain
\begin{equation}
m_1 \; \approx \; m_2 \; \approx \; 7.7 \times 10^{-2} ~ {\rm eV} \; .
\end{equation}
We see that the neutrino mass spectra of Patterns $\rm B_1$ and $\rm B_2$ have
much similarity too.

\vspace{0.3cm}
 
\underline{Pattern $\rm B_3$:} ~ To lowest order,
the neutrino mass spectrum in this pattern is identical to that in Pattern 
$\rm B_1$. 

\vspace{0.3cm}

\underline{Pattern $\rm B_4$:} ~ To lowest order, 
the neutrino mass spectrum in this pattern is identical to that in
Pattern $\rm B_2$. We have seen that the phenomenological consequences 
of Patterns $\rm B_1$, $\rm B_2$, $\rm B_3$ and $\rm B_4$ are 
nearly the same \cite{Xing02,FGM}. It is particularly difficult to 
distinguish between Patterns $\rm B_1$ and $\rm B_3$,
or between Patterns $\rm B_2$ and $\rm B_4$. 

\vspace{0.3cm}

\underline{Pattern C:} ~ In the lowest-order approximation, we obtain
\begin{equation}
m_3 \; \approx \; |t_{2y}| s_z 
\sqrt{\frac{\Delta m^2_{\rm atm}}{t_x | t_x 
+ 2 t_{2y} s_z c_\delta |}} \;\; .
\end{equation}
One can see that the magnitude of $m_3$ depends not only upon three
mixing angles $(\theta_x, \theta_y, \theta_z)$ but also upon the
Dirac phase $\delta$. Therefore a careful analysis of the parameter
space is required for Pattern C \cite{Y}, 
so as to fit current experimental data and
to fix the allowed range of $m_3$. The values of $\delta$
and $\theta_z$ are unfortunately unknown at present. Just for the
purpose of illustration, we typically take $\theta_x = \theta_y = 44.8^\circ$, 
$\theta_z = 5^\circ$ and $\delta = 90^\circ$. In this special case,
one may get $R_\nu \approx 0.03$ -- a correct order of
$\Delta m^2_{\rm sun}/\Delta m^2_{\rm atm}$ \cite{Xing02}. 
Then we arrive at
\begin{equation}
m_3 \; \approx \; \frac{t_{2y} s_z}{t_x} \sqrt{\Delta m^2_{\rm atm}} 
\; \approx \; 0.7 ~ {\rm eV} \; .
\end{equation}
Using the same inputs, we get
$m_1 \approx m_2 \approx m_3 \approx 0.7 ~ {\rm eV}$ to a high degree of
accuracy. This result indicates that three neutrino masses are
essentially degenerate, and their magnitude can be of ${\cal O}(1)$ eV.
Therefore it is rather sensitive to the neutrinoless double beta decay.

\vspace{0.5cm}

Indeed two-zero textures of the neutrino mass matrix allow us to obtain 
definite predictions for the neutrinoless double beta decay, whose 
effective mass term is a simple function of neutrino masses and flavor mixing
parameters:
\begin{equation}
\langle m\rangle_{ee} \; =\; m_3 \left | V^2_{e1} \xi + V^2_{e2} \zeta 
+ V^2_{e3} \right | \; .
\end{equation}
Using the parametrization of $V$ in Eq. (2) and the
expressions of $\xi$ and $\zeta$ in Table 1, we
can calculate $\langle m\rangle_{ee}$ for each of the seven patterns of $M$.
The approximate analytical results are listed in Table 2. Some
comments are in order.

\vspace{0.3cm}

(1) $\langle m\rangle_{ee} \approx 0$ holds for Patterns $\rm A_1$ and
$\rm A_2$. This is obviously true, as $M_{ee} =0$ has been taken in
both patterns.

\vspace{0.3cm}

(2) The sizes of $\langle m\rangle_{ee}$ in Patterns $\rm B_1$ and $\rm B_3$
are essentially identical:
$\langle m\rangle_{ee} \approx m_1 \approx m_2 \approx 5.4 \times 10^{-2}$ eV
for $\theta_y = 40^\circ$. 
So are the sizes of $\langle m\rangle_{ee}$ in 
Patterns $\rm B_2$ and $\rm B_4$:
$\langle m\rangle_{ee} \approx m_1 \approx m_2 \approx 7.7 \times 10^{-2}$ eV
for $\theta_y = 40^\circ$. 

\vspace{0.3cm}

(3) If $\theta_x = \theta_y = 44.8^\circ$, $\theta_z = 5^\circ$ and
$\delta = 90^\circ$ are typically taken, one will arrive at
$\langle m\rangle_{ee} \approx m_1 \approx m_2 \approx m_3 \approx 0.7$ eV
for Pattern C
\footnote{At this point we notice that there is a typing error associated
with $|M_{ee}|$ in Eq. (29) of
Ref. \cite{Xing02}. The correct result should be
$|M_{ee}| \approx m_3 \sqrt{1 - 4c_\delta/(t_{2x} t_{2y} s_z)
+ 4/(t^2_{2x} t^2_{2y} s^2_z)} ~$.}.
This result is apparently consistent with the alleged evidence for the
neutrinoless double beta decay \cite{Beta}, 
$0.05$ eV $\leq \langle m\rangle_{ee} \leq$ 0.84 eV, at the $95\%$
confidence level. Nevertheless, it might be premature to take this
measurement too seriously \cite{Beta2}. To be more conservative,
we take account of the relatively reliable experimental upper bound 
$\langle m\rangle_{ee} < 0.35$ eV 
(at the $90\%$ confidence level \cite{Beta3}) and find that
the typical result $\langle m\rangle_{ee} = 0.7$ eV is actually disfavored.
It should be noted, however, that there does exist the appropriate parameter
space for Pattern C \cite{FGM}, in which both 
$R_\nu \sim {\cal O}(10^{-2})$ and 
$\langle m\rangle_{ee} < 0.35$ eV can be satisfied.

\vspace{0.3cm}

As pointed out in Ref. \cite{FGM}, the seven patterns of $M$ can be
classified into three distinct categories: A (with $\rm A_1$ and
$\rm A_2$), B (with $\rm B_1$, $\rm B_2$, $\rm B_3$ and $\rm B_4$),
and C. A definite measurement of the neutrinoless double beta decay 
at the level $\langle m\rangle_{ee} \sim {\cal O} (0.01)$ eV to 
${\cal O} (0.1)$ eV will rule out category A. On the other hand,
more precise data of neutrino oscillations will help reduce the
uncertainties associated with $\langle m\rangle_{ee}$
predicted in categories B and C. It might even be possible to
distinguish between categories B and C (and therefore discard
one of them), if the magnitude of $\langle m\rangle_{ee}$ is 
experimentally determined in the future.

\vspace{0.5cm}

In summary, we have shown that a full determination of the neutrino
mass spectrum is indeed possible from two-zero textures of the
neutrino mass matrix $M$. This important observation indicates that
two-zero textures of $M$ have much more predictability than previously
expected. In particular, one can get definite predictions for
the neutrinoless double beta decay.
We hope that a variety of proposed precision measurements of 
neutrino oscillations and lepton-number-violating processes may 
finally allow us to pin down the unique texture of lepton mass
matrices.

\vspace{0.5cm}

The author would like to thank H.J. He, D. Marfatia and T. Yoshikawa
for useful discussions. He is also grateful to H.V. Klapdor-Kleingrothaus
for pointing out an error in the list of references. 
This work was supported in part by the National Natural Science 
Foundation of China.

\newpage

\begin{table}
\caption{Seven patterns of the neutrino mass matrix $M$ with two independent
vanishing entries, which were found to be in accord with current 
experimental data and empirical hypotheses \cite{Xing02,FGM}. 
Analytical results for the neutrino mass ratios
$\xi\equiv m_1/m_3$ and $\zeta\equiv m_2/m_3$ are given in terms of the 
flavor mixing parameters $\theta_x$, $\theta_y$, $\theta_z$ and 
$\delta$ \cite{Xing02}, where $t_x \equiv \tan\theta_x$,
$t_{2y} \equiv \tan 2\theta_y$, $s_z \equiv \sin\theta_z$,
$s_{2x} \equiv \sin 2\theta_x$, $c_\delta \equiv \cos\delta$ and so on.}
\begin{center}
\begin{tabular}{ccccl} \hline\hline 
Pattern &~~& Texture of $M$ &~~~& Results of $\xi$ and $\zeta$
\\ \hline 
$\rm A_1$ 
&& $\left ( \matrix{
{\bf 0} & {\bf 0} & \times \cr
{\bf 0} & \times & \times \cr
\times & \times & \times \cr} \right )$
&& 
$\displaystyle \xi \approx t_x t_y s_z \; , ~~~
\zeta \approx \frac{t_y}{t_x} s_z$
\\ \hline 
$\rm A_2$ 
&& $\left ( \matrix{
{\bf 0} & \times & {\bf 0} \cr
\times & \times & \times \cr
{\bf 0} & \times & \times \cr} \right )$
&& 
$\displaystyle \xi \approx \frac{t_x}{t_y} s_z \; , ~~~
\zeta \approx \frac{1}{t_x t_y} s_z$
\\ \hline
$\rm B_1$ 
&& $\left ( \matrix{
\times & \times & {\bf 0} \cr
\times & {\bf 0} & \times \cr
{\bf 0} & \times & \times \cr} \right )$
&& 
$\displaystyle \xi \approx 
\zeta \approx t^2_y \; , ~~~
\xi - \zeta \approx 
\frac{4s_z c_\delta}{s_{2x} s_{2y}}$
\\ \hline 
$\rm B_2$ 
&& $\left ( \matrix{
\times & {\bf 0} & \times \cr
{\bf 0} & \times & \times \cr
\times & \times & {\bf 0} \cr} \right )$
&& 
$\displaystyle \xi \approx 
\zeta \approx \frac{1}{t^2_y} \; , ~~~
\zeta - \xi \approx 
\frac{4s_z c_\delta}{s_{2x} s_{2y}}$
\\ \hline
$\rm B_3$ 
&& $\left ( \matrix{
\times & {\bf 0} & \times \cr
{\bf 0} & {\bf 0} & \times \cr
\times & \times & \times \cr} \right )$
&& 
$\displaystyle \xi \approx 
\zeta \approx t^2_y \; , ~~~
\zeta - \xi \approx 
\frac{4 t^2_y s_z c_\delta}{s_{2x} s_{2y}}$
\\ \hline 
$\rm B_4$ 
&& $\left ( \matrix{
\times & \times & {\bf 0} \cr
\times & \times & \times \cr
{\bf 0} & \times & {\bf 0} \cr} \right )$
&& 
$\displaystyle \xi \approx 
\zeta \approx \frac{1}{t^2_y} \; , ~~~
\xi - \zeta \approx 
\frac{4 s_z c_\delta}{s_{2x} s_{2y} t^2_y}$
\\ \hline
$\rm C$ 
&& $\displaystyle \left ( \matrix{
\times & \times & \times \cr
\times & {\bf 0} & \times \cr
\times & \times & {\bf 0} \cr} \right )$
&& 
$\displaystyle \xi \approx 
\sqrt{1 - \frac{2c_\delta}{t_x t_{2y} s_z} + 
\frac{1}{t^2_x t^2_{2y} s^2_z}} \;\; , ~~~
\zeta \approx 
\sqrt{1 + \frac{2 t_x c_\delta}{t_{2y} s_z} + 
\frac{t^2_x}{t^2_{2y} s^2_z}}$
\\ \hline\hline
\end{tabular}
\end{center}
\end{table}

\begin{table}
\caption{Seven patterns of the neutrino mass matrix $M$ with two independent
vanishing entries, and their predictions for $\langle m\rangle_{ee}$ of the 
neutrinoless double beta decay, in which $t_x \equiv \tan\theta_x$,
$t_{2y} \equiv \tan 2\theta_y$, $s_z \equiv \sin\theta_z$,
$c_\delta \equiv \cos\delta$ and so on.}
\begin{center}
\begin{tabular}{ccccl} \hline\hline 
Pattern &~~& Texture of $M$ &~~~& Result of $\langle m\rangle_{ee}$
\\ \hline 
$\rm A_1$ 
&& $\left ( \matrix{
{\bf 0} & {\bf 0} & \times \cr
{\bf 0} & \times & \times \cr
\times & \times & \times \cr} \right )$
&& 
$0$
\\ \hline 
$\rm A_2$ 
&& $\left ( \matrix{
{\bf 0} & \times & {\bf 0} \cr
\times & \times & \times \cr
{\bf 0} & \times & \times \cr} \right )$
&& 
$0$
\\ \hline
$\rm B_1$ 
&& $\left ( \matrix{
\times & \times & {\bf 0} \cr
\times & {\bf 0} & \times \cr
{\bf 0} & \times & \times \cr} \right )$
&& 
$\displaystyle 
t^2_y \sqrt{\frac{\Delta m^2_{\rm atm}}{|t^4_y -1|}}$
\\ \hline 
$\rm B_2$ 
&& $\left ( \matrix{
\times & {\bf 0} & \times \cr
{\bf 0} & \times & \times \cr
\times & \times & {\bf 0} \cr} \right )$
&& 
$\displaystyle 
\sqrt{\frac{\Delta m^2_{\rm atm}}{|t^4_y -1|}}$
\\ \hline
$\rm B_3$ 
&& $\left ( \matrix{
\times & {\bf 0} & \times \cr
{\bf 0} & {\bf 0} & \times \cr
\times & \times & \times \cr} \right )$
&& 
$\displaystyle 
t^2_y \sqrt{\frac{\Delta m^2_{\rm atm}}{|t^4_y -1|}}$
\\ \hline 
$\rm B_4$ 
&& $\left ( \matrix{
\times & \times & {\bf 0} \cr
\times & \times & \times \cr
{\bf 0} & \times & {\bf 0} \cr} \right )$
&& 
$\displaystyle 
\sqrt{\frac{\Delta m^2_{\rm atm}}{|t^4_y -1|}}$
\\ \hline
$\rm C$ 
&& $\displaystyle \left ( \matrix{
\times & \times & \times \cr
\times & {\bf 0} & \times \cr
\times & \times & {\bf 0} \cr} \right )$
&& 
$\displaystyle \frac{|t_{2y}|}{t_x} s_z
\sqrt{\frac{\Delta m^2_{\rm atm}}{|t_x + 2 t_{2y} s_z c_\delta|}}
\sqrt{1 - \frac{4c_\delta}{t_{2x} t_{2y} s_z} + 
\frac{4}{t^2_{2x} t^2_{2y} s^2_z}}$
\\ \hline\hline
\end{tabular}
\end{center}
\end{table}
\normalsize


\newpage
\begin{thebibliography}{99}
\bibitem{Xing02} Z.Z. Xing, Phys. Lett. B {\bf 530} (2002) 159.

\bibitem{ATM} Super-Kamiokande Collaboration, Y. Fukuda {\it et al.}, 
Phys. Lett. B {\bf 467} (1999) 185;
S. Fukuda {\it et al.}, Phys. Rev. Lett. {\bf 85} (2000) 3999.

\bibitem{SUN} Super-Kamiokande Collaboration, S. Fukuda {\it et al.}, 
Phys. Rev. Lett. {\bf 86} (2001) 5651; 
Phys. Rev. Lett. {\bf 86} (2001) 5656;
SNO Collaboration, Q.R. Ahmad {\it et al.}, 
Phys. Rev. Lett {\bf 87} (2001) 071301;
nucl-ex/0204008; nucl-ex/0204008.

\bibitem{CHOOZ} CHOOZ Collaboration, M. Apollonio {\it et al.},
Phys. Lett. B {\bf 420} (1998) 397; 
Palo Verde Collaboration, F. Boehm {\it et al.},
Phys. Rev. Lett. {\bf 84} (2000) 3764.

\bibitem{FGM} P.H. Frampton, S.L. Glashow, and D. Marfatia,
hep-ph/0201008, to appear in Phys. Lett. B.

\bibitem{FOY} P.H. Frampton, M.C. Oh, and T. Yoshikawa,
hep-ph/0204273.

\bibitem{Tanimoto} A. Kageyama, S. Kaneko, Shimoyama, and M. Tanimoto,
hep-ph/0204291.

\bibitem{FX01} H. Fritzsch and Z.Z. Xing, Phys. Lett. B {\bf 517} (2001) 363;
Z.Z. Xing, Phys. Rev. D {\bf 64} (2001) 073014.

\bibitem{X02} Z.Z. Xing, Phys. Rev. D {\bf 65} (2002) 077302;
hep-ph/0204050.

\bibitem{Bahcall} V. Barger, D. Marfatia, K. Whisnant, and B.P. Wood,
hep-ph/0204253; 
J.N. Bahcall, M.C. Gonzalez-Garcia, and 
C. Pe$\rm\tilde{n}$a-Garay, hep-ph/0204314.

\bibitem{FX96} See, e.g., 
H. Fritzsch and Z.Z. Xing, Phys. Lett. B {\bf 372} (1996);
Phys. Lett. B {\bf 440} (1998) 313; 
Phys. Rev. D {\bf 61} (2000) 073016;
M. Fukugida, M. Tanimoto, and T. Yanagida, Phys. Rev. D {\bf 57} (1998) 4429;
K. Kang and S.K. Kang, hep-ph/9802328;
H.J. He, D.A. Dicus, and J.N. Ng, Phys. Lett. B {\bf 536} (2002) 83;
J.I. Silva-Marcos, hep-ph/0204051.
For a review with more extensive references, see:
H. Fritzsch and Z.Z. Xing, 
Prog. Part. Nucl. Phys. {\bf 45} (2000) 1.

\bibitem{Yanagida} M. Fujii, K. Hamaguchi, and T. Yanagida,
hep-ph/0202210;
W. Buchm$\rm\ddot{u}$ller, hep-ph/0204288.

\bibitem{PDG} Particle Data Group, D.E. Groom {\it et al.},
Eur. Phys. J. C {\bf 15} (2000) 1.

\bibitem{Y} T. Yoshikawa, private communications.

\bibitem{Beta} H.V. Klapdor-Kleingrothaus, A. Dietz, H.L. Harney,
and I.V. Krivosheina, Mod. Phys. Lett. A {\bf 16} (2001) 2409.

\bibitem{Beta2} S. Pascoli and S.T. Petcov, hep-ph/0205022;
C.E. Aalseth {\it et al.}, hep-ex/0202018; 
H.V. Klapdor-Kleingrothaus, hep-ex/0205228.

\bibitem{Beta3} Heidelberg-Moscow Collaboration, H.V. Klapdor-Kleingrothaus,
hep-ph/0103074; C.E. Aalseth {\it et al.}, hep-ex/0202026.

\end{thebibliography}
\end{document}